\documentclass[12pt,prd,nofootinbib,showkeys]{revtex4}

\usepackage{bm}
\usepackage{graphics}
\usepackage{rotating}
\usepackage{epsfig}
\usepackage{amsmath}
\usepackage{amsfonts}
\usepackage{amssymb}

\usepackage{multirow}

\newcommand{\be}{\begin{equation}}
\newcommand{\ee}{\end{equation}}
\newcommand{\bdis}{\begin{displaymath}}
\newcommand{\edis}{\end{displaymath}}
\newcommand{\bga}{\begin{equation}\begin{gathered}}
\newcommand{\ega}{\end{gathered}\end{equation}}

\begin{document}
\title{Pair ${(bc)}$ diquarks production in high energy\\ proton--proton collisions}
\author{\firstname{Anton} \surname{Trunin}}
\affiliation{Bogoliubov Laboratory of Theoretical Physics, Joint Institute for Nuclear Research, Joliot--Curie Street 6, 141980 Dubna, Russia}
\affiliation{Samara State Aerospace University, Moskovskoye Shosse 34, 443086 Samara, Russia}
\email{amtrnn@gmail.com}

\begin{abstract}

The cross section of pair double heavy diquark production process $pp\to(bc)+(\bar b \bar c)+X$  is calculated in the leading order of gluonic fusion channel with all four possible color and spin combinations $[^1S_0]_{\bar3}$, $[^1S_0]_{6}$, $[^3S_1]_{\bar3}$, and $[^3S_1]_{6}$ for each of the two final diquarks taken into account. Several sources of relativistic corrections to the cross section are handled in the framework of relativistic quark model. 
Perturbative $\mathcal O(v^2)$ corrections originating from the production amplitude expansions in heavy quark relative velocity~$v$ depend on the color and spin states of the final particles, but can be generally considered as unimportant ones giving maximally 12\% improvement in numerically significant cases. Modifications of the quark--quark and antiquark--antiquark bound state wave functions caused by the appropriate generalization of the Breit interaction potential have rather severe impact on the cross section suppressing it almost three times.
Under assumption of antitriplets and sextuplets' nonperturbative parameters having the same order of magnitude, it is shown that the color-sextet mechanism strongly dominates pair diquark production in both nonrelativistic and relativistic approximations.

\end{abstract}



\maketitle

\section{Introduction}

At the moment, contrary to the broadly studied charmonium and single heavy baryon families, there are no concrete experimental results on baryons containing two heavy $b$ or $c$ quarks. Initially, their first experimental observation was reported by SELEX collaboration more than 10 years ago, but it is still not confirmed in any subsequent experimental study followed since then~\cite{selex,baryons2}. Theoretical investigation of the double heavy baryons motivated by the uncertain experimental situation as well as by unique properties and wide range of physics associated with such systems represents an actively developing field~\cite{likh2002,bramb05,chang06,albertus,chang,karliner14,cheng15,karliner,zheng16}.
The double heavy baryons provide a remarkable opportunity to test quantum chromodynamics~(QCD) and several effective theories based on it in both hard and low-energy regions.
The heavy quarks $Q$ and $Q'$ within the baryon $(QQ'q)$ are predicted to form a heavy diquark~--- compact quark--quark bound state $(QQ')$ in  antitriplet or, alternatively, sextuplet color state~\cite{fleck88,diquark}.
With respect to the picture of strong interaction a heavy diquark in antitriplet color state is equal to an antiquark in heavy-light meson: the smallness of diquark radius $r_{QQ'}\ll\Lambda_\text{QCD}$ allows to consider it as almost static and point-like source of gluonic field, so that the dynamics of the light quark $q$ in both types of hadrons is expected to be quite similar.
Therefore, the double heavy baryons combine the aspects of both heavy--heavy and heavy--light quark bound states and can serve as an independent test object for the respective models and theoretical constructions, like NRQCD~\cite{nrqcd} and HQET~\cite{hqet}.
Moreover, there is close connection between the double heavy baryons and even more unusual diquark bound states, such as double heavy tetraquarks~\cite{karliner,tetra}. The latter can be considered as possible candidates for some states from the broad list of exotic ``XYZ'' resonances discovered during the last years~\cite{exotics}.

According to the quark--diquark model, the production of double heavy baryon is divided in two stages. On the first step, which is described by perturbative QCD, the creation of two quark--antiquark pairs $Q\bar Q$ and $Q'\bar Q'$ takes place.
On the second step, the created quarks and antiquarks rearrange to form the bound state of heavy diquark $(QQ')$ or $(\bar Q\bar Q')$ with its subsequent hadronization to the observable double heavy baryon. The transition of heavy diquark into the baryon is generally covered by the appropriate fragmentation functions $D_{(QQ')\to(QQ'q)}(z)$~\cite{likh2002,fragm}. Nevertheless, taking into account the several order of magnitude difference between heavy diquark and light quark masses, the diquark can be assumed to carry almost all of the final baryon momentum, so that the corresponding fragmentation function approximates to $D_{(QQ')\to(QQ'q)}(z)=\delta(1-z)P_{(QQ')\to(QQ'q)}$~\cite{diquark}. 
Applicability of such approximation was confirmed by direct calculations in Ref.~\cite{chen14-2}.
Although the heavy diquark $(QQ')$ bounds the light quark $q$ very easily, there is also a possibility for the diquark dissociation events decreasing the total transition probability $P_{(QQ')\to(QQ'q)}<1$. Therefore, cross sections of double heavy diquarks represent an upper bound for the yield of double heavy baryons in the same reaction summed over light quark flavors and over all possible baryon spin states. The nonperturbative stage of heavy baryon production can be described by the appropriate matrix elements of NRQCD, for which, due to the lack of experimental data, the potential model predictions for $|\Psi_{(QQ')}(0)|^2$ are used.

The pair production of double heavy baryons in $e^+e^-$ annihilation and $pp$--collisions was studied in Refs.~\cite{brag02,mt2014,dq2014}.
In all these studies, however, only the case of (anti)triplet color states of two final diquarks has been considered.
As it was already mentioned, the heavy diquark can be produced either in an antitriplet or in a sextuplet color configuration representing a close analogy to the color-singlet and color-octet mechanisms of heavy quarkonium production~\cite{nrqcd}. Both octet and sextuplet cases require an additional gluon to be emitted in the nonperturbative part of the process, what generally $\mathcal O(v^2)$ suppresses the appropriate matrix elements, if the required emission is attributed to the heavy quark of relative velocity $v$. Nevertheless, in  the case of double heavy baryon the final state also contains a light quark, which produces gluon easily, so that the different power counting rules can be applied, and both antitriplet and sextuplet matrix elements turn out to be of the same order~\cite{ma}. Under this assumption, the sextuplet mechanism was shown to be equally or even more important than conventional antitriplet channel for various high energy processes of double heavy baryon production~\cite{ma,chang07,zhang11,jiang12,jiang13,chen14,chen14-2}.

This paper is focused on the pair production of double heavy $(bc)$ diquarks in proton--proton collisions with
all four possible color and spin combinations $[^1S_0]_{\bar3}$, $[^1S_0]_{6}$, $[^3S_1]_{\bar3}$, $[^3S_1]_{6}$ for each of the two final diquarks taken into account. The calculation technique is based on the notion of relativistic quark model with the elements of quasipotential approach~\cite{quasi}, which first application to the problems of pair charmonium production was demonstrated in Ref.~\cite{ebert1}.
Our consideration is limited by the gluon fusion processes, for which the initial state of proton collision is approximated by gluonic pair $gg$. The alternative possibilities, such as so-called ``intrinsic charm'' processes with initial states $gc$ or $cc$, were studied in Refs.~\cite{chang06,chang07} in connection with single $\Xi_{cc}$ baryon production. There it was shown, that intrinsic charm mechanism is crucial under SELEX kinematical conditions, while its contributions are much less important in Tevatron and LHC cases.
The paper is organized as follows: in Section~\ref{section2} the general formalism of the applied approach is briefly described. The exact expression for the relativistic production amplitude is given in the leading order in strong coupling constant, and the general formulae for the cross sections of pair diquark production are presented with the account of second order relativistic corrections in $v$. Section~\ref{section3} contains numerical details of the model and results for the cross sections calculated at the LHC c.m. energies $\sqrt S=7$ and 14~TeV in nonrelativistic and relativistic approximations. The role of sextuplet contributions is discussed, and several sources of relativistic corrections are analyzed.
The Appendix delivers a short note on the structure of electronic supplementary material to the paper.

\section{General formalism}
\label{section2}
The cross section of pair double heavy diquark production in proton--proton collisions can be presented in the following form corresponding to the collinear approximation for colliding protons~\cite{dq2014,likh89}:
\be
\label{eq:cs-plus-x}
d\sigma[p+p\to D_{bc} + \bar D_{\bar b \bar c}+X]=\int \! dx_1 dx_2 \, f_{g/p}(x_1,\mu) f_{g/p}(x_2,\mu)
\, d\sigma[gg\to D_{bc} + \bar D_{\bar b \bar c}],
\ee
where $f_{g/p}(x,\mu)$ is the partonic distribution function for gluon in proton, $x_{1,2}$ are the
longitudinal momentum fractions of gluons, $\mu$ is the factorization scale.
Neglecting the proton mass and taking
c.m. reference frame of the initial protons with the beam along the $z$-axis
we can present the gluon on mass-shell momenta as $k_{1,2}=x_{1,2}\frac{\sqrt{S}}{2}(1,0,0,\pm 1)$. At the high center-of-mass energy $\sqrt{S}$ in proton--proton collisions, the main contribution to the cross section~\eqref{eq:cs-plus-x} is expected to come from the gluon fusion process $gg\to D_{bc} + \bar D_{\bar b \bar c}$.

Taking into account two spin and color states for each of the final diquarks, there are 16 different sub-processes contributing to the pair diquark production. Only 10 of them are independent, while cross sections for the rest can be obtained from the respective symmetry relations. Nevertheless, contributions from all 16 processes have eventually to be summed up in order to obtain an estimate for pair double heavy baryon production.
In the quasipotential approach the production amplitude for the gluonic
sub-process $gg\to D_{bc} + \bar D_{\bar b \bar c}$ can be expressed as a convolution of the perturbative production amplitude of $(bc)$ and $(\bar b \bar c)$ (anti)quark pairs $\mathcal T(p_1,p_2;q_1,q_2)$ and the
quasipotential wave functions of the final diquarks $\Psi_{(bc)}(p,P)$ and $\Psi_{(\bar b\bar c)}(q,Q)$~\cite{ebert1}:
\be
\label{eq:m-gen}
{\mathcal M}[gg\to D_{bc} + \bar D_{\bar b \bar c}](k_1,k_2,P,Q)=\int \! \frac{d\mathbf p}{(2\pi)^3}
\int \! \frac{d\mathbf q}{(2\pi)^3} \, \bar\Psi_{(bc)}(p,P) \bar\Psi_{(\bar b\bar c)}(q,Q) \otimes \mathcal T(p_1,p_2;q_1,q_2),
\ee
where $p_{1,2}$ are four-momenta of $c$ and $b$ quarks, and $q_{1,2}$ are the appropriate four-momenta for $\bar c$ and $\bar b$ antiquarks. They are defined in terms of total momenta $P(Q)$ and relative momenta $p(q)$ as follows:
\bga
p_{1,2}=\eta_{1,2} P \pm p,\quad (pP)=0, \qquad q_{1,2}=\rho_{1,2} Q \pm q,\quad (qQ)=0, \\
\eta_{1,2} = \frac{M_{bc}^2 \pm m_c^2 \mp m_b^2}{2M_{bc}^2}, \qquad \rho_{1,2} = \frac{M_{\bar b\bar c}^2 \pm m_c^2 \mp m_b^2}{2M_{\bar b\bar c}^2},
\ega
where $m_{c,b}$ are quark masses, $M_{bc}=M_{D_{bc}}$ and $M_{\bar b\bar c}=M_{\bar D_{\bar b \bar c}}$ are diquark masses, $p=L_P(0,\mathbf p)$ and $q=L_Q(0,\mathbf q)$ are the relative four-momenta obtained by the Lorentz transformation of four-vectors
$(0,\mathbf p)$ and $(0,\mathbf q)$ to the reference frames moving with the four-momenta $P$~and~$Q$ of the final diquarks $D_{bc}$ and $\bar D_{\bar b \bar c}$, respectively. The integration in Eq.~\eqref{eq:m-gen} is performed over the relative three-momenta of quarks and antiquarks
in the final state.

\begin{figure}[t!]
\includegraphics[scale=0.75]{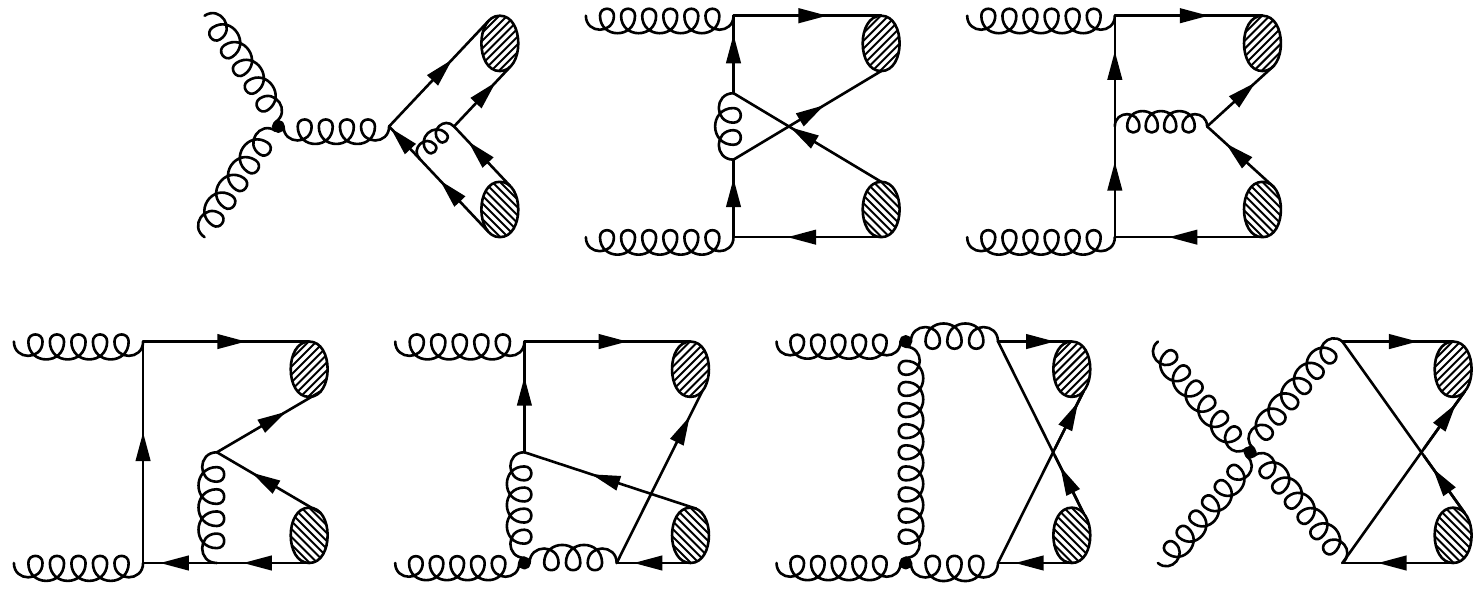}
\caption{The leading order diagrams contributing to  
 $gg\to D_{bc} + \bar D_{\bar b \bar c}$.
The others can be obtained by reversing the quark lines or interchanging the initial gluons.
}
\label{fig:d35}
\end{figure}

\begin{figure}[t!]
\includegraphics[scale=0.75]{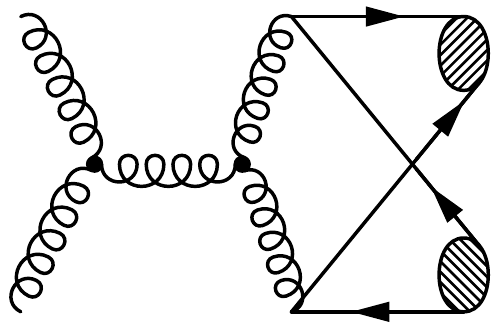}
\caption{One additional leading order diagram contributing only to the sub-processes with color states $\bar3+\bar6$ or $6+3$ of the final diquark pair $D_{bc} + \bar D_{\bar b \bar c}$.
}
\label{fig:d36}
\end{figure}

In the leading order in the strong coupling constant $\alpha_s$, there are 36 Feynman
diagrams describing the process \hbox{$gg\to D_{bc} + \bar D_{\bar b \bar c}$}, which are presented in Fig.~\ref{fig:d35} and Fig.~\ref{fig:d36}.
The single diagram from Fig.~\ref{fig:d36} produces non-zero contribution only if the resulting pair $D_{bc} + \bar D_{\bar b \bar c}$ contains diquarks with the color combination $\bar3+\bar6$ or $6+3$, and it still always vanishes for the final state consisting of two scalar diquarks $SD_{bc} + S\bar D_{\bar b \bar c}$.
Due to the large volume of calculations the package FeynArts~\cite{feynarts}
for Mathematica was used to obtain analytical expressions for all diagrams, and their traces were subsequently calculated with Form~\cite{form}.
Then the leading order production amplitude~\eqref{eq:m-gen} reads 
 \bdis
\mathcal M[gg\to D_{bc} + \bar D_{\bar b \bar c}](k_1,k_2,P,Q)=\sqrt{M_{bc}\,M_{\bar b\bar c}}\,\pi^2\alpha_s^2\int\!\frac{d\mathbf p}{(2\pi)^3} \int\!\frac{d\mathbf q}{(2\pi)^3} \mathrm{Tr}\,\mathfrak M,
 \edis 
 \bdis
\mathfrak M= \bar\Psi_{P,p}^{bc}\gamma_\beta\bar\Psi_{Q,q}^{cb}\gamma_\omega\Gamma_1^{\beta\omega}+\bar\Psi_{P,-p}^{cb}\gamma_\beta\,\Gamma_2^{\beta\omega\theta} \gamma_\omega\bar\Psi_{Q,-q}^{bc}\gamma_\theta+\bar\Psi_{P,p}^{bc}\gamma_\beta\,\Gamma_3^{\beta\omega\theta} \gamma_\omega\bar\Psi_{Q,q}^{cb}\gamma_\theta 
 \edis 
 \bdis +
\bar\Psi_{P,p}^{bc}\,\hat\varepsilon_1\frac{m_c-\hat k_1+\hat p_1}{(k_1-p_1)^2-m_c^2}\gamma_\beta(\bar\Psi_{Q,q}^{cb}\gamma_\omega\Gamma_4^{\beta\omega}+\Gamma_5^{\beta\omega}\bar\Psi_{Q,q}^{cb}\gamma_\omega)
\edis
\bdis
+ \bar\Psi_{P,-p}^{cb}\,\hat\varepsilon_1\frac{m_b-\hat k_1+\hat p_2}{(k_1-p_2)^2-m_b^2}\gamma_\beta(\bar\Psi_{Q,-q}^{bc}\gamma_\omega\Gamma_6^{\beta\omega}+\Gamma_7^{\beta\omega}\bar\Psi_{Q,-q}^{bc}\gamma_\omega) 
 \edis 
 \bdis
+\bar\Psi_{P,p}^{bc}\,\hat\varepsilon_2\frac{m_c-\hat k_2+\hat p_1}{(k_2-p_1)^2-m_c^2}\gamma_\beta\Gamma_8^{\beta\omega}\bar\Psi_{Q,q}^{cb}\gamma_\omega
+\bar\Psi_{P,-p}^{cb}\,\hat\varepsilon_2\frac{m_b-\hat k_2+\hat p_2}{(k_2-p_2)^2-m_b^2}\gamma_\beta\Gamma_9^{\beta\omega}\bar\Psi_{Q,-q}^{bc}\gamma_\omega
 \edis 
\begin{equation}
\label{eq:m-main}
+\bar\Psi_{P,-p}^{cb}\gamma_\beta\frac{m_b+\hat k_1-\hat q_2}{(k_1-q_2)^2-m_b^2}\hat\varepsilon_1\bar\Psi_{Q,-q}^{bc}\gamma_\omega\Gamma_{10}^{\beta\omega}
+\bar\Psi_{P,p}^{bc}\gamma_\beta\frac{m_c+\hat k_1-\hat q_1}{(k_1-q_1)^2-m_c^2}\hat\varepsilon_1\bar\Psi_{Q,q}^{cb}\gamma_\omega\Gamma_{11}^{\beta\omega},
\end{equation}
where $\varepsilon_{1,2}$ are polarization vectors of the initial gluons, the hat symbol means contraction of the four-vector with the Dirac gamma-matrices, and vertex functions $\Gamma_i$ were introduced  to make the entry of the amplitude~\eqref{eq:m-main} more compact. The normalization factors $\sqrt{2M_{bc}}$ and $\sqrt{2M_{\bar b\bar c}}$ of the quasipotential bound state wave functions were explicitly extracted in~\eqref{eq:m-main}.

The formation of diquark states from (anti)quark
pairs, which corresponds to the first stage of the double heavy baryon formation, is determined in the quark model by the quasipotential wave functions $\Psi_{bc}(p,P)$ and
$\Psi_{\bar b \bar c}(q,Q)$. These wave functions are calculated initially
in the diquark rest frame and then transformed to the reference frames moving with the four-momenta
$P$ and $Q$. The law of such transformation was derived in the Bethe--Salpeter approach in Ref.~\cite{brodsky2}
and in the quasipotential method in Ref.~\cite{faustov}. The last one gives the following expressions for the relativistic wave functions~\cite{mt2014}:
\begin{equation}
\begin{gathered}
\label{eq:relwf}
\bar\Psi_{P,p}^{bc}=\frac{\bar\Psi_{bc}^{0}(\mathbf p)}{\sqrt{ \frac{e_c(p)}{m_c} \frac{e_c(p)+m_c}{2m_c} \frac{e_b(p)}{m_b} \frac{e_b(p)+m_b}{2m_b} }}
\left[
	\frac{\hat v_1-1}{2}+\hat v_1\frac{\mathbf p^2}{2m_b(e_b(p)+m_b)}-\frac{\hat p}{2m_b}
\right] \\ \times
\Sigma^P(1+\hat v_1)
\left[
	\frac{\hat v_1+1}{2}+\hat v_1\frac{\mathbf p^2}{2m_c(e_c(p)+m_c)}+\frac{\hat p}{2m_c}
\right],
\\
\bar\Psi_{Q,q}^{cb}=\frac{\bar\Psi_{\bar b \bar c}^{0}(\mathbf q)}{\sqrt{ \frac{e_c(q)}{m_c} \frac{e_c(q)+m_c}{2m_c} \frac{e_b(q)}{m_b} \frac{e_b(q)+m_b}{2m_b} }}
\left[
	\frac{\hat v_2-1}{2}+\hat v_2\frac{\mathbf q^2}{2m_c(e_c(q)+m_c)}+\frac{\hat q}{2m_c}
\right] \\ \times
\Sigma^Q(1+\hat v_2)
\left[
	\frac{\hat v_2+1}{2}+\hat v_2\frac{\mathbf q^2}{2m_b(e_b(q)+m_b)}-\frac{\hat q}{2m_b}
\right],
\end{gathered}
\end{equation}
where $e_{c,b}(p)=\sqrt{p^2+m_{c,b}^2}$, $v_1 = P/M_{bc}$, $v_2 = Q/M_{\bar b \bar c}$, and $\Sigma^{P,Q}$ equal to $\gamma_5$ and $\hat\varepsilon_{P,Q}$ for scalar ($S$) and axial-vector ($AV$) diquarks, respectively. The polarization vectors $\varepsilon_{P,Q}$ of the axial-vector diquarks fulfill the relations $(\varepsilon_{P}P) = 0$ and $(\varepsilon_{Q}Q) = 0$. The quasipotential wave functions~\eqref{eq:relwf} include (anti)quark pairs projection operators for the given spin states $\bar u_i(0)\bar u_j(0)=[C\hat\varepsilon(\gamma_5)(1+\gamma_0)]_{ij}/2\sqrt2$ and $v_i(0)v_j(0)=[(1-\gamma_0)\hat\varepsilon(\gamma_5)C]_{ij}/2\sqrt2$, $C$ is the charge conjugation matrix. Note that the order of the upper indices $b$ and $c$ in the left hand side of definitions~\eqref{eq:relwf} is important, so that their permutation leads to the proper replacements $m_c \leftrightarrow m_b$ apllied to the r.h.s. of Eqs.~\eqref{eq:relwf}.

Leading order vertex functions $\Gamma_i$ in~\eqref{eq:m-main} have the following explicit form:
\bdis
\Gamma_1^{\beta\omega}=
{\mathcal K_1} D_{\mu}{}^{\beta}(p_1+q_1)D_{\nu}{}^{\omega}(p_2+q_2)\bigl((2-\kappa_c)(\varepsilon_1^\nu \varepsilon_2^\mu+\kappa_c\varepsilon_1^\mu \varepsilon_2^\nu)-(1+\kappa_c)g^{\mu\nu}(\varepsilon_1 \varepsilon_2)
 \edis 
 \bdis
-iD_{\lambda\kappa}(k_1-p_1-q_1)\mathfrak{E}_1^{\lambda\mu}(p_1+q_1)\mathfrak{E}_2^{\kappa\nu}(p_2+q_2) -i\kappa_c  
D_{\kappa\lambda}(k_1-p_2-q_2)\mathfrak{E}_1^{\kappa\nu}(p_2+q_2)\mathfrak{E}_2^{\lambda\mu}(p_1+q_1)
 \edis 
 \bdis
 +i(1-\kappa_c) D_{\kappa\theta}(k_1+k_2)
 \mathfrak{E}_1^{\kappa}(-k_2)\mathfrak{L}^{\mu\theta\nu}(k_1+k_2,p_2+q_2)\bigr),
 \edis 
 \bdis
\Gamma_2^{\beta\omega\theta}=\kappa_s{\mathcal K_{2}}\mathfrak{E}_2^{\mu}(-k_1) D_{\mu}{}^{\beta}(k_1+k_2)D^{\theta\omega}(p_1+q_1)\frac{m_b-\hat p_1-\hat q_1-\hat q_2}{(p_1+q_1+q_2)^2-m_b^2} 
+\kappa_s
{\mathcal K_{12}}\varepsilon_2^\omega \mathfrak{E}_1^{\mu\nu}(p_1+q_1)
 \edis 
 \bdis 
 \times
 D_{\mu}{}^{\beta}(k_1-p_1-q_1)D_{\nu}{}^{\theta}(p_1+q_1) \frac{m_b+k_2-q_2}{(k_2-q_2)^2-m_b^2} 
+\kappa_s\Delta_c
D^{\theta\beta}(p_1+q_1)\frac{m_b+\hat p_1+\hat p_2+\hat q_1}{(p_1+p_2+q_1)^2-m_b^2}
 \edis 
 \bdis 
 \times
 \Bigr(
 \kappa_c
{\mathcal K_{2}}\mathfrak{E}_1^{\mu}(-k_2) D_{\mu}{}^{\omega}(k_1+k_2)
+{\mathcal K_{9}}\varepsilon_1^\omega\hat\varepsilon_2\frac{m_b+\hat k_1-\hat q_2}{(k_1-q_2)^2-m_b^2} 
+{\mathcal K_{7}}\varepsilon_2^\omega\hat\varepsilon_1\frac{m_b+\hat k_2-\hat q_2}{(k_2-q_2)^2-m_b^2} \Bigr), 
 \edis 
 \bdis
\Gamma_4^{\beta\omega}={\mathcal K_3}D^{\beta\omega}(k_1-p_1-q_1)\frac{m_b+\hat k_2-\hat p_2}{(k_2-p_2)^2-m_b^2}\hat\varepsilon_2 - 
{\mathcal K_4}\varepsilon_2^\omega D^{\beta\mu}(k_1-p_1-q_1)\frac{m_b-\hat k_2+\hat q_2}{(k_2-q_2)^2-m_b^2}\gamma_\mu 
 \edis 
 \bdis -
{\mathcal K_5}\mathfrak{E}_2^{\mu\nu}(p_2+q_2)D_{\mu}{}^{\beta}(k_1-p_1-q_1)D_{\nu}{}^{\omega}(p_2+q_2) ,
 \edis 
 \bdis
\Gamma_5^{\beta\omega}={\mathcal K_6}D^{\beta\omega}(p_2+q_2)\frac{m_c+\hat k_2-\hat q_1}{(k_2-q_1)^2-m_c^2}\hat\varepsilon_2 + 
{\mathcal K_7}\varepsilon_2^\beta D_{\mu}{}^{\omega}(p_2+q_2)\frac{m_c-\hat p_2-\hat q_1-\hat q_2}{(p_2+q_1+q_2)^2-m_c^2}\gamma_\mu,
 \edis 
 \bdis
\Gamma_8^{\beta\omega}={\mathcal K_8}D^{\beta\omega}(p_2+q_2)\frac{m_c+\hat k_1-\hat q_1}{(k_1-q_1)^2-m_c^2}\hat\varepsilon_1
+ {\mathcal K_9}\varepsilon_1^\beta D_{\mu}{}^{\omega}(p_2+q_2)\frac{m_c-\hat p_2-\hat q_1-\hat q_2}{(p_2+q_1+q_2)^2-m_c^2}\gamma_\mu 
 \edis 
 \bdis
-{\mathcal K_{10}}\mathfrak{E}_1^{\mu\nu}(p_2+q_2)D_{\mu}{}^{\beta}(k_1-p_2-q_2)D_{\nu}{}^{\omega}(p_2+q_2),
 \edis
\bdis
\Gamma_{10}^{\beta\omega}=\kappa_s{\mathcal K_{11}}D^{\beta\omega}(k_1-p_2-q_2)\frac{m_c+\hat k_2-\hat p_1}{(k_2-p_1)^2-m_c^2}\hat\varepsilon_2
+\kappa_s\Delta_c{\mathcal K_{3}}\varepsilon_2^\omega D^{\beta\mu}(k_1-p_2-q_2)\frac{m_c-\hat k_2+\hat q_1}{(k_2-q_1)^2-m_c^2}\gamma_\mu
\edis
\begin{equation}
\label{eq:gammas}
+\kappa_s{\mathcal K_{13}}\mathfrak{E}_2^{\mu\nu}(p_1+q_1)D_{\mu}{}^{\beta}(k_1-p_2-q_2)D_{\nu}{}^{\omega}(p_1+q_1),
\end{equation}
where the constant $\kappa_s$ equals $+1$ in the case of the equal spin states  of the final diquark pair $S(AV)+S(AV)$ and $-1$ for the other two possibilities $S(AV)+AV(S)$. Analogously, $\kappa_c=+1$ for $\bar 3(6)+3(\bar6)$, $\kappa_c=-1$ for $\bar 3(6)+\bar6(3)$, and the values of $\Delta_c=1/2,\,2,\,1$ correspond to the final color states $6+3$, $\bar3+\bar6$, and $\bar 3(6)+3(\bar6)$, respectively. The following tensors are introduced:
\begin{equation}
\begin{gathered}
\mathfrak{L}^{\mu\nu\omega}(x,y)=g^{\mu\nu}(y^\omega-2x^\omega)-g^{\omega\mu}(2y^\nu-x^\nu)+g^{\omega\nu}(x^\mu+y^\mu),\\
\mathfrak{E}_{1,2}^{\mu\nu}(x)=\varepsilon_{1,2}^\omega\mathfrak{{L}^{\mu\nu}}_{\omega}(x,k_{1,2})
, \qquad \mathfrak{E}_{1,2}^{\mu}(x)=\varepsilon_{2,1}^\nu\mathfrak{E}_{1,2}^{\mu\nu}(x),
\end{gathered}
\end{equation}
and $D_{\mu\nu}(k)$ is the gluon propagator, which is subsequently taken in the Feynman gauge $D^{\mu\nu}_\text{F}(k)=-ig^{\mu\nu}/k^2$.
The additional vertex functions $\Gamma_i$ can be found by simultaneous replacement $m_c \leftrightarrow m_b$, $p_1 \leftrightarrow p_2$, and $q_1 \leftrightarrow q_2$ in Eqs.~\eqref{eq:gammas}:
\begin{equation}
\begin{gathered}
\label{eq:gammas2}
\Gamma_3^{\beta\omega\theta}=\kappa_s\kappa_c\Gamma_2^{\beta\omega\theta}\biggl|{}_{\substack{m_b \rightleftharpoons m_c \\ p_1 \rightleftharpoons p_2 \\ q_1 \rightleftharpoons q_2}}, \quad
\Gamma_6^{\beta\omega}=\kappa_s\kappa_c\Gamma_4^{\beta\omega}\biggl|{}_{\substack{m_b \rightleftharpoons m_c \\ p_1 \rightleftharpoons p_2 \\ q_1 \rightleftharpoons q_2}}, \quad
\Gamma_7^{\beta\omega}=\kappa_s\kappa_c\Gamma_5^{\beta\omega}\biggl|{}_{\substack{m_b \rightleftharpoons m_c \\ p_1 \rightleftharpoons p_2 \\ q_1 \rightleftharpoons q_2}}, \\
\Gamma_9^{\beta\omega}=\kappa_s\kappa_c\Gamma_8^{\beta\omega}\biggl|{}_{\substack{m_b \rightleftharpoons m_c \\ p_1 \rightleftharpoons p_2 \\ q_1 \rightleftharpoons q_2}}, \quad
\Gamma_{11}^{\beta\omega}=\kappa_s\kappa_c\Gamma_{10}^{\beta\omega}\biggl|{}_{\substack{m_b \rightleftharpoons m_c \\ p_1 \rightleftharpoons p_2 \\ q_1 \rightleftharpoons q_2}}.
\end{gathered}
\end{equation}

The 36 leading order Feynman diagrams summed with the antisymmetric color functions $\epsilon^{c_1c_2A}/\sqrt2$ ($c_i,A = 1,2,3$) of (anti)triplet diquarks and/or with the partially symmetric functions $d^{c_1c_2A}/\sqrt2$ ($A = 1,\ldots,6$) of (anti)sextuplets lead to the nontrivial color structure of the production amplitude, which manifests itself through the color factors $\mathcal K_i$ in~\eqref{eq:gammas}.
For the $\bar 3+3$ final state we have%
\footnote{The author is grateful to S. P.~Baranov for pointing out that color structure of the amplitude can be additionally simplified in this case.}:
\bdis
{\mathcal K_1}=-2{\mathcal C_0}-3{\mathcal C_1}+6{\mathcal C_2}, \quad {\mathcal K_2}=\frac43{\mathcal C_1}, \quad
{\mathcal K_3}=\frac{2i}3({\mathcal C_0}+2{\mathcal C_1}-4{\mathcal C_2}), 
\quad
{\mathcal K_4}=\frac{i}3({\mathcal C_0}-{\mathcal C_1}-{\mathcal C_2}), 
 \edis
 \bdis
{\mathcal K_5}={\mathcal C_0}+{\mathcal C_1}-3{\mathcal C_2}, \quad
{\mathcal K_6}=-\frac{i}{3}({\mathcal C_0}+3{\mathcal C_1}-5{\mathcal C_2}), 
\quad
{\mathcal K_7}=\frac{2i}{3}({\mathcal C_0}-2{\mathcal C_2}), \quad
{\mathcal K_8}=-\frac{i}3({\mathcal C_0}+2{\mathcal C_1}-5{\mathcal C_2}), 
 \edis 
 \bdis
{\mathcal K_9}=\frac{2i}3({\mathcal C_0}+2{\mathcal C_1}-2{\mathcal C_2}),  \quad
{\mathcal K_{10}}={\mathcal C_0}+2{\mathcal C_1}-3{\mathcal C_2},  \quad
{\mathcal K_{11}}=-\frac{i}3({\mathcal C_0}+2{\mathcal C_1}-{\mathcal C_2}), \quad
{\mathcal K_{12}}=\mathcal K_{5}, 
 \edis
\be
\label{eq:ks1}
{\mathcal K_{13}}=\mathcal K_{10}, \qquad
\mathcal C_0=\delta^{g_1g_2}\delta_{AB}, \quad
\mathcal C_1=if^{g_1g_2a}(T^a)_{BA}, \quad
\mathcal C_2=(T^{g_1}T^{g_2})_{BA}.
\ee
For $\bar3+\bar6$:
\bdis
{\mathcal K_1}={\mathcal C_0}, \quad 
{\mathcal K_2}=-\frac23{\mathcal C_0}, \quad
{\mathcal K_3}=\frac{2i}3{\mathcal C_1}, \quad
{\mathcal K_4}=\frac{i}3({\mathcal C_1}-3{\mathcal C_2}), \quad
{\mathcal K_5}={\mathcal C_1}-{\mathcal C_2}, \quad
{\mathcal K_6}=-\frac{i}{3}(3{\mathcal C_1}-{\mathcal C_2}), 
 \edis 
 \bdis
{\mathcal K_7}=-\frac{2i}{3}{\mathcal C_2}, \quad
{\mathcal K_8}=-\frac{i}3({\mathcal C_0}-3{\mathcal C_1}-{\mathcal C_2}), \quad
{\mathcal K_9}=\frac{2i}3({\mathcal C_0}-{\mathcal C_2}),  \quad
{\mathcal K_{10}}={\mathcal C_0}-{\mathcal C_1}-{\mathcal C_2},
 \edis
\be
\label{eq:ks2}
{\mathcal K_{11}}=-\frac{i}3(3{\mathcal C_0}-{\mathcal C_1}-3{\mathcal C_2}),\quad
{\mathcal K_{12}}=-{\mathcal C_1}-{\mathcal C_2},\quad
{\mathcal K_{13}}={\mathcal C_0}+{\mathcal C_1}-{\mathcal C_2}.
\ee
For $6+3$:
\bdis
{\mathcal K_1}=-{\mathcal C_0}, \quad 
{\mathcal K_2}=\frac43{\mathcal C_0}, \quad
{\mathcal K_3}=-\frac{4i}3{\mathcal C_1}, \quad
{\mathcal K_4}=\frac{i}3({\mathcal C_1}+3{\mathcal C_2}), \quad
{\mathcal K_5}=-{\mathcal C_1}+{\mathcal C_2}, \quad
{\mathcal K_6}=\frac{i}{3}(3{\mathcal C_1}+{\mathcal C_2}), 
 \edis 
 \bdis
{\mathcal K_7}=\frac{4i}{3}{\mathcal C_2}, \quad
{\mathcal K_8}=-\frac{i}3({\mathcal C_0}+3{\mathcal C_1}-{\mathcal C_2}), \quad
{\mathcal K_9}=-\frac{4i}3({\mathcal C_0}-{\mathcal C_2}),  \quad
{\mathcal K_{10}}=-{\mathcal C_0}+{\mathcal C_1}+{\mathcal C_2},
 \edis
\be
\label{eq:ks3}
{\mathcal K_{11}}=\frac{i}3(3{\mathcal C_0}+{\mathcal C_1}-3{\mathcal C_2}),\quad
{\mathcal K_{12}}={\mathcal C_1}+{\mathcal C_2},\quad
{\mathcal K_{13}}=-{\mathcal C_0}-{\mathcal C_1}+{\mathcal C_2}.
\ee
And finally, for $6+\bar6$:
 \bdis
{\mathcal K_1}=-{\mathcal C_0}-2{\mathcal C_1}+2{\mathcal C_2}, \quad 
{\mathcal K_2}=\frac23{\mathcal C_0}, \quad
{\mathcal K_3}=\frac{2i}3{\mathcal C_1}, \quad
{\mathcal K_4}=\frac{i}3({\mathcal C_1}-3{\mathcal C_2}), \quad
{\mathcal K_5}={\mathcal C_1}-{\mathcal C_2},
 \edis 
 \bdis
{\mathcal K_6}=-\frac{i}{3}(3{\mathcal C_1}-{\mathcal C_2}),\quad
{\mathcal K_7}=-\frac{2i}{3}{\mathcal C_2}, \quad
{\mathcal K_8}=-\frac{i}3({\mathcal C_0}+3{\mathcal C_1}-{\mathcal C_2}), \quad
{\mathcal K_9}=\frac{2i}3({\mathcal C_0}-{\mathcal C_2}),
 \edis
\be
\label{eq:ks4}
{\mathcal K_{10}}={\mathcal C_0}+{\mathcal C_1}-{\mathcal C_2}, \quad
{\mathcal K_{11}}=-\frac{i}3(3{\mathcal C_0}+{\mathcal C_1}-3{\mathcal C_2}),\quad
{\mathcal K_{12}}={\mathcal K_{5}},\quad
{\mathcal K_{13}}={\mathcal K_{10}}.
\ee
The coefficients ${\mathcal C}_i$ are written explicitly in Eq.~\eqref{eq:ks1} for the $\bar 3+3$ final color state, where $f^{ijk}$ and $T^a$ are $SU(3)$ structure constants and fundamental representation generators, $g_{1,2} = 1, \ldots, 8$ are color indices of initial gluons, $A$ and $B$ are color indices of the  diquarks $D_{bc}$ and $\bar D_{\bar b \bar c}$, respectively.
Note that the color factors of Ref.~\cite{dq2014} agree with Eq.~\eqref{eq:ks1} after the equality $\mathcal C_3=f^{g_1ca}f^{g_2cb}(T^aT^b)_{BA}=\mathcal C_0/4+3\,\mathcal C_2/2$ applied.
For all other color combinations the following general form is valid:
 \bdis
\mathcal C_0=if^{g_1g_2a}(T^a)_{c_1c_3}c_f^{c_1c_2A}c_f^{c_3c_2B}, \quad
\mathcal C_1=(T^{g_1})_{c_1c_3}(T^{g_2})_{c_2c_4}c_f^{c_1c_2A}c_f^{c_3c_4B}, 
 \edis
\be
\label{eq:cfs}
\mathcal C_2=(T^{g_1}T^{g_2})_{c_1c_3}c_f^{c_1c_2A}c_f^{c_3c_2B},
\ee
where $c_f$ stands for the corresponding color wave function $\epsilon$ or $d$.

The production amplitude~\eqref{eq:m-main} and vertex
functions~\eqref{eq:gammas} contain relative momenta $p$ and $q$ in exact form. In order to take
into account relativistic corrections of the second order in $p$ and $q$ the expansion is performed for all
inverse denominators of the quark and gluon propagators:
\begin{equation}
\begin{gathered}
\label{eq:exps}
\frac{1}{(p_{1,2}+q_{1,2})^2}=\frac{1}{Z_0}\Bigl[
1 \mp \frac{2(\eta_{1,2}\,pQ+\rho_{1,2}\,qP)}{Z_0}-\frac{p^2+2pq+q^2}{Z_0} + \ldots \Bigr], \\
\frac{1}{(p_1+q_1+q_2)^2-m_b^2} = 
\frac{1}{Z_1} \Bigl[ 1 - \frac{2pQ+p^2}{Z_1}+\frac{4(pQ)^2}{Z_1^2}  + \ldots \Bigr], \\
\frac{1}{(k_2-q_1)^2-m_c^2} = \frac{1}{Z_2} \Bigl[ 1 + \frac{2k_2q-q^2}{Z_2}+\frac{4(k_2q)^2}{Z_2^2}  + \ldots \Bigr],
\end{gathered}
\end{equation}
where  $s=(k_1+k_2)^2=(P+Q)^2=x_1x_2S$ and $t=(P-k_1)^2=(Q-k_2)^2$ are the Mandelstam variables for the gluonic sub-process, and leading order expansion denominators are $Z_0=s\,\eta_{1,2}\rho_{1,2}+(\eta_{1,2}-\rho_{1,2})(\eta_{1,2}M_{bc}^2-\rho_{1,2}M_{\bar b\bar c}^2)$, $Z_1 = s\,\eta_1-\eta_2(\eta_1 M_{bc}^2-M_{\bar b\bar c}^2) - m_b^2$, and $Z_2 = t\,\rho_1-\rho_1\rho_2 M_{\bar b\bar c}^2 - m_c^2$. The amplitude~\eqref{eq:m-main} contains 16 different denominators to be expanded in the manner of Eq.~\eqref{eq:exps}. Temporarily neglecting the bound state corrections, it can be found that expansion denominators have one of the following form: $s\,\eta_{1,2}$, $s\,\eta_{1,2}^2$, $\eta_{1,2}(M^2-t)$ or $\eta_{1,2}(M^2-s-t)$. Then, taking into account kinematical restrictions for $s$ and $t$, along with the nonrelativistic estimate \hbox{$\eta_1=\rho_1= m_c/(m_c+m_b) \approx 1/4$} for $(bc)$ diquarks, one can conclude that expansion parameters in~\eqref{eq:exps} are at least as small as $4p^2/M^2$ and $4q^2/M^2$.
Preserving in the expanded amplitude terms up to the second order both in the relative momenta $p$ and $q$,
the angular integration can be performed according to the relations for $S$-wave states:
\bdis
\int\!\frac{\Psi_0(\mathbf p)}{\sqrt{ \frac{e_c(p)}{m_c} \frac{e_c(p)+m_c}{2m_c} \frac{e_b(p)}{m_b} \frac{e_b(p)+m_b}{2m_b} }}\frac{d\mathbf p}{(2\pi)^3}=\frac{1}{\sqrt{2}\,\pi}\int\limits_0^\infty\!\frac{p^2 R_p(p)}{\sqrt{ \frac{e_c(p)}{m_c} \frac{e_c(p)+m_c}{2m_c} \frac{e_b(p)}{m_b} \frac{e_b(p)+m_b}{2m_b} }}dp,
\edis
\be
\int\! \frac{p_\mu p_\nu \, \Psi_0(\mathbf p)}{\sqrt{ \frac{e_c(p)}{m_c} \frac{e_c(p)+m_c}{2m_c} \frac{e_b(p)}{m_b} \frac{e_b(p)+m_b}{2m_b} }}\frac{d\mathbf p}{(2\pi)^3}=-\frac{g_{\mu\nu}-{v_1}_\mu{v_1}_\nu}{3\sqrt2\,\pi}
\int\limits_0^\infty\!\frac{p^4 R_p(p)}{\sqrt{ \frac{e_c(p)}{m_c} \frac{e_c(p)+m_c}{2m_c} \frac{e_b(p)}{m_b} \frac{e_b(p)+m_b}{2m_b} }}dp,
\ee
where $R_p(p)$ is the radial wave function in the momentum representation.

In order to calculate the cross section the squared modulus of the amplitude have to be summed upon the final polarizations in the case of axial-vector diquarks and also to be averaged over polarizations of the initial gluons using the following relations:
\be
\sum_{\lambda}\varepsilon_{P,Q}^\mu \, {\varepsilon_{P,Q}^\ast}^\nu = v_{1,2}^\mu v_{1,2}^\nu-g^{\mu\nu},\qquad
\sum_{\lambda}\varepsilon_{1,2}^\mu \, {\varepsilon_{1,2}^\ast}^\nu=\frac{k_1^\mu k_2^\nu+k_1^\nu k_2^\mu}{(k_1k_2)}-g^{\mu\nu}.
\ee
Then it is also averaged over $8\times8$ possible initial gluons color states and summed over diquarks color indices $A$ and $B$:
\be
\label{eq:color-sum}
\epsilon^{c_1c_2A}\epsilon^{c_3c_4A}=\delta^{c_1c_3}\delta^{c_2c_4}-\delta^{c_1c_4}\delta^{c_2c_3},\qquad
d^{c_1c_2A}d^{c_3c_4A}=\delta^{c_1c_3}\delta^{c_2c_4}+\delta^{c_1c_4}\delta^{c_2c_3}.
\ee
In the case of the diquark pair with identical spin states and masses the cross section can be presented as 
\bga
\label{eq:cs}
d\sigma[gg\to D_{bc} + \bar D_{\bar b \bar c}](s,t)=\frac{\pi M_{bc}M_{\bar b\bar c}\, \alpha_s^4}{65\,536\,s^2}| R(0)|^4  \bigl[ F^{(1)}(s,t) -3(2\omega_{01}+2\omega_{10}-\omega_{11}) F^{(1)}(s,t) 
\\+\frac{27}{2}(\omega_{01}+\omega_{10})^2 F^{(1)}(s,t) + \frac12\omega_{\frac12\frac12}(2-9\omega_{01}-9\omega_{10})F^{(2)}(s,t)
+\omega_{\frac12\frac12}^2 F^{(3)}(s,t) \bigr],
\ega
while for the final state containing particles of unequal masses it has more complicated form:
\bga
\label{eq:cs2}
d\sigma[gg\to S(AV)D_{bc} + AV(S)\bar D_{\bar b \bar c}](s,t)=\frac{\pi M_{bc}M_{\bar b\bar c}\, \alpha_s^4}{65\,536\,s^2}| R_{SD_{bc}}(0)|^2| R_{AV\bar D_{\bar b \bar c}}(0)|^2 \\
\times  \bigl[ F^{(1)}(s,t) -3(\omega^S_{01}+\omega^S_{10}+\omega^{AV}_{01}+\omega^{AV}_{10}-\frac12\omega^S_{11}-\frac12\omega^{AV}_{11}) F^{(1)}(s,t) 
+9(\omega^S_{01}+\omega^S_{10})(\omega^{AV}_{01}+\omega^{AV}_{10})
\\ \times F^{(1)}(s,t)
+\frac94(\omega^S_{01}+\omega^S_{10})^2F^{(1)}(s,t)+\frac94(\omega^{AV}_{01}+\omega^{AV}_{10})^2F^{(1)}(s,t) 
+ \frac14\omega^S_{\frac12\frac12}(2-3\omega^S_{01}-3\omega^S_{10}\\-6\omega^{AV}_{01}-6\omega^{AV}_{10})F^{(2)}(s,t)
+ \frac14\omega^{AV}_{\frac12\frac12}(2-3\omega^{AV}_{01}-3\omega^{AV}_{10}-6\omega^S_{01}-6\omega^S_{10})F^{(2)}(s,t)
\\
+\omega^S_{\frac12\frac12}\omega^{AV}_{\frac12\frac12} F^{(3)}(s,t)
+\frac14(\omega^S_{\frac12\frac12}+\omega^{AV}_{\frac12\frac12})^2 F^{(4)}(s,t) \bigr].
\ega
The relativistic parameters $\omega_{nk}$ are expressed through momentum integrals of the double heavy diquark radial wave function $R_p(p)$:
\bga
\label{eq:is}
I_{nk} = \int_0^{m_c} \!\!\! p^2 R_p(p)\sqrt{\frac{(e_c(p)+m_c)(e_b(p)+m_b)}{2e_c(p) \, 2e_b(p)}} 
\biggl( \frac{e_c(p)-m_c}{e_c(p)+m_c} \biggr)^n
\biggl( \frac{e_b(p)-m_b}{e_b(p)+m_b} \biggr)^k
dp
\\
\omega_{nk} = \sqrt\frac{2}{\pi}\frac{I_{nk}}{R(0)},
\ega
and
\bga
\label{eq:r0def}
R(0)=\sqrt\frac{2}{\pi}\int_0^{\infty} \!\!\! p^2R_p(p)\,dp
\ega
is the value of coordinate wave function at the origin.
The first term in Eqs.~\eqref{eq:cs} and~\eqref{eq:cs2} is proportional to the single function $F^{(1)}(s,t)$ and corresponds to the zeroth order result in heavy quark velocity expansion of the cross sections, and the other terms containing products of the form $\omega_{nk}F^{(i)}(s,t)$
represent relativistic corrections to it. The analytical results for the functions $F^{(i)}(s,t)$ are too lengthy and cannot be given here in an explicit form, but they are provided as a supplementary material to the electronic version of this paper.

\section{Numerical results and discussion}
\label{section3}
The quasipotential wave functions of the double heavy diquarks in (anti)triplet color state are obtained by numerical solving of the Schr\"odinger equation with effective relativistic Hamiltonian based on the QCD generalization of the Breit potential, which was additionally improved by the scalar and vector exchange confinement terms as it is described in details in Ref.~\cite{mt2014}. The values of diquark masses and relativistic parameters defined by Eq.~\eqref{eq:is} are given in Table~\ref{tbl0}. There is no such model available for sextuplet diquarks, so all their nonperturbative parameters, including $ R(0)$, are simply taken equal to the ones of triplet diquarks, what is the common practice allowing to roughly estimate the order of contributions from these color states~\cite{chang06,ma,chang07,zhang11,jiang12,jiang13,chen14,chen14-2}. The relativistic parameters $\omega_{nk}$ from Table~\ref{tbl0} are almost indistinguishable for scalar and axial-vector $(bc)$ diquarks, and this fact was used to slightly simplify the final view of the cross section~\eqref{eq:cs2}.
The definition~\eqref{eq:is} of the relativistic integrals $I_{nk}$ contains a cutoff at the value of $c$-quark mass $\Lambda = m_c$. Although the integrals~\eqref{eq:is} are numerically convergent, the relativistic wave function cannot be reliably calculated in the region $p\gtrsim m_c$ in the considered model. It should also be noted that the formulation of effective relativistic Hamiltonian in Ref.~\cite{mt2014} assumes expansion of the interaction potential in powers of ${|\mathbf p|}/\frac{2m_cm_b}{m_b+m_c}\approx \frac23|\mathbf p|/m_c$ and the expansion parameters of Eqs.~\eqref{eq:exps} were estimated as $\sim\frac14|\mathbf p|^2/m_c^2$.
So, the current choice of cutoff for relativistic terms preserves the validity of both relativistic expansions and also prevents the contribution of large numerical errors connected with multiplication of the wave function by additional $p^2$ factors in the highly relativistic region.
There is no such numerical instability involved in calculation of the parameter $R(0)$ determining zeroth order contribution in $v$, which is therefore defined over the whole available momentum range in both nonrelativistic and relativistic approximations. The nonrelativistic version of this parameter $R^\text{NR}(0)$ has the same formal definition $R^\text{NR}(0)=\sqrt{2/\pi}\int p^2R^\text{NR}_p(p)\,dp$ with the only difference that it is calculated with the simple Cornell potential model $V^\text{NR}(r)=-2\alpha_s/(3r)+1/2(Ar+B)$ in contrast to the complicated Breit-based Hamiltonian used to obtain $R(0)$. Therefore, the parameter $R(0)$ purely reflects the nonperturbative relativistic effects of quark--quark interaction within the bound state, while all auxiliary normalization factors of the wave function transformation law~\eqref{eq:relwf} have been incorporated into the correction terms $\omega_{nk}F^{(i)}(s,t)$ in Eqs.~\eqref{eq:cs} and~\eqref{eq:cs2}.

\begin{table}[t!]
\caption{\label{tbl0}Numerical values of the parameters describing scalar and axial-vector $(bc)$ diquarks.}
\bigskip
\begin{tabular}{lcccccccc}
\hline
Diquark & $n^{2S+1}L_J$ & $M$, & $R^\text{NR}(0)$ & $ R(0)$, & $\omega_{10}$ &$\omega_{01}$ & $\omega_{\frac12\frac12}$   &
$\omega_{11}$  \\
state  &     &   GeV  & GeV$^{3/2}$ & GeV$^{3/2}$  &    &     &    &    \\    \hline
$SD_{bc}$	& $1^1S_0$	& 6.517 & 0.67 & 0.54 & 0.0383 & 0.0045 & 0.0131  & 0.00039  \\ %
$AVD_{bc}$  & $1^3S_1$  & 6.526 & 0.67 & 0.52 & 0.0384 & 0.0045 & 0.0132  & 0.00038  \\  \hline
\end{tabular}
\end{table}

\begin{table}[t!]
\caption{\label{tbl1}Cross sections of the pair $(bc)$ diquarks production in proton--proton collisions (nb). The results in nonrelativistic approximation (nonrel.) and with relativistic effects (rel.) are given.}

\begin{tabular}{lcc|cc|cc|cc}
\hline
& \multicolumn{4}{c}{$\sqrt S=7$~TeV} & \multicolumn{4}{c}{$\sqrt S=14$~TeV} \\
Diquark pair & \multicolumn{2}{c}{CTEQ5L}  & \multicolumn{2}{c|}{CTEQ6L1} & \multicolumn{2}{c}{CTEQ5L}  & \multicolumn{2}{c}{CTEQ6L1} \\ 
 & $\sigma_{nonrel.}$ & $\sigma_{rel.}$ & $\sigma_{nonrel.}$ & $\sigma_{rel.}$ & $\sigma_{nonrel.}$ & $\sigma_{rel.}$ & $\sigma_{nonrel.}$ & $\sigma_{rel.}$ \\ \hline
$S{(bc)}_{\bar3}+S{(\bar b\bar c)}_{3}$ & 0.063 & 0.023 & 0.057 & 0.021 & 0.14 & 0.05 & 0.12 & 0.043\\
$S{(bc)}_{\bar3}+S{(\bar b\bar c)}_{\bar 6}$ & 0.007 & 0.003 & 0.007 & 0.003 & 0.016 & 0.006 & 0.014 & 0.005\\
$S{(bc)}_{6}+S{(\bar b\bar c)}_{\bar 6}$ & 0.04 & 0.015 & 0.037 & 0.013 & 0.088 & 0.032 & 0.078 & 0.029\\[5pt]
$AV{(bc)}_{\bar3}+AV{(\bar b\bar c)}_{3}$ & 0.25 & 0.068 & 0.23 & 0.062 & 0.55 & 0.15 & 0.48 & 0.13\\
$AV{(bc)}_{\bar3}+AV{(\bar b\bar c)}_{\bar 6}$ & 0.29 & 0.076 & 0.27 & 0.07 & 0.63 & 0.17 & 0.56 & 0.15\\
$AV{(bc)}_{6}+AV{(\bar b\bar c)}_{\bar 6}$ & 1.4 & 0.38 & 1.3 & 0.35 & 3.1 & 0.84 & 2.7 & 0.74\\[5pt]
$S{(bc)}_{\bar3}+AV{(\bar b\bar c)}_{3}$ & 0.031 & 0.009 & 0.029 & 0.009 & 0.069 & 0.02 & 0.061 & 0.018\\
$S{(bc)}_{\bar3}+AV{(\bar b\bar c)}_{\bar 6}$ & 0.17 & 0.051 & 0.15 & 0.047 & 0.37 & 0.11 & 0.32 & 0.098\\
$S{(bc)}_{6}+AV{(\bar b\bar c)}_{3}$ & 0.16 & 0.047 & 0.14 & 0.043 & 0.34 & 0.1 & 0.3 & 0.091\\
$S{(bc)}_{6}+AV{(\bar b\bar c)}_{\bar 6}$ & 0.14 & 0.041 & 0.12 & 0.037 & 0.3 & 0.089 & 0.26 & 0.078\\
\hline
\end{tabular}

\end{table}

The numerical results for the total cross section of pair scalar and axial-vector $(bc)$ diquarks production corresponding to the LHC energies $\sqrt S=7$ and 14~TeV are presented in Table~\ref{tbl1}. The integration in~\eqref{eq:cs-plus-x} was performed with the partonic distribution functions CTEQ5L and CTEQ6L1~\cite{cteqs}. Both renormalization and factorization scales were set to the $\mu=\sqrt{(M_{bc}+M_{\bar b\bar c})^2/4+P_T^2}$, where $P_T$ is the transverse momentum of the final diquarks, and leading order result for the strong coupling with the initial condition $\alpha_s(M_Z)=0.118$ was used.
In the nonrelativistic limit all parameters $\omega_{nk}$ are taken to be equal zero, so that only $F^{(1)}(s,t)$ term survives in square brackets of Eq.~\eqref{eq:cs}, and diquark masses are assumed equal to the sum of its constituent (anti)quark masses $M_0=m_b + m_c$. 
The cross sections calculated in such approximation are marked as $\sigma_{nonrel}$ in Table~\ref{tbl1}.
As it shown in Table~\ref{tbl0}, the nonrelativistic parameter $R^\text{NR}(0)$ has the value $R^\text{NR}(0)=0.67$~GeV$^{3/2}$ for $(bc)$ diquarks in the considered model~\cite{mt2014,dq2014} lying close to the result $0.73$~GeV$^{3/2}$ from Ref.~\cite{gershtein}.

\begin{figure}[t!]
\vspace*{.7cm}
\begin{center}
\includegraphics{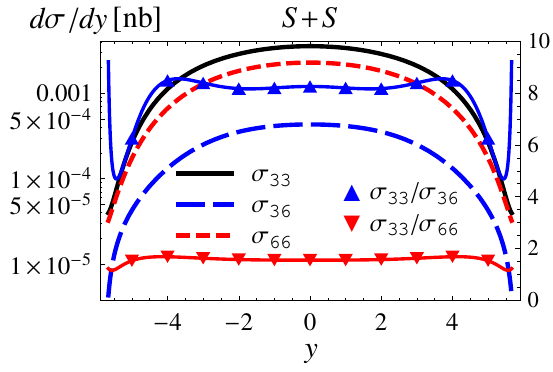}\includegraphics{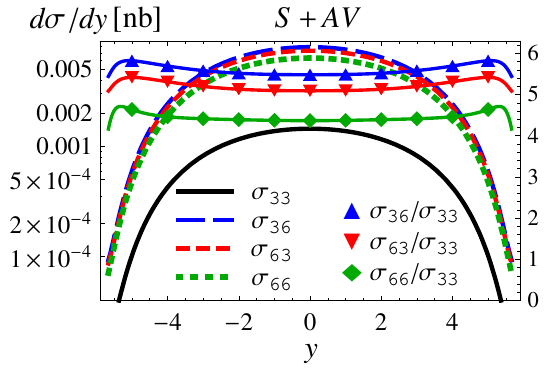}\includegraphics{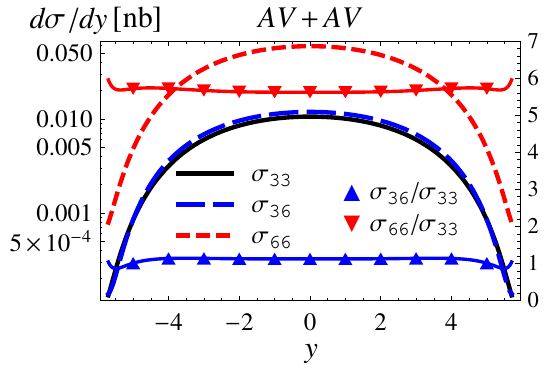}
\end{center}
\vspace{-.7cm}
\caption{The rapidity $y\equiv y_{D_{bc}}$ distributions for $S+S$ (left), $S+AV$ (center), and $AV+AV$ (right) pair double heavy diquark production at $pp$--collision energy $\sqrt S=7$~TeV. The scale at the right edge of figures corresponds to the ratios of cross sections. The compact notations $\sigma_{33}=d\sigma[(bc)_{\bar3}+(\bar b\bar c)_3]/dy$ and so on are used in the figures.}
\label{fig:y-distr}
\end{figure}

\begin{figure}[t!]
\vspace*{.7cm}
\begin{center}
\includegraphics{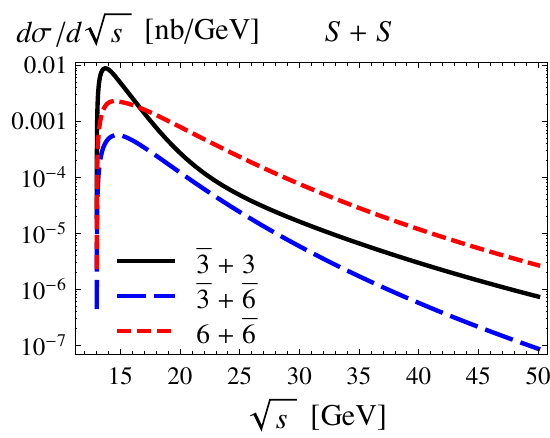}\includegraphics{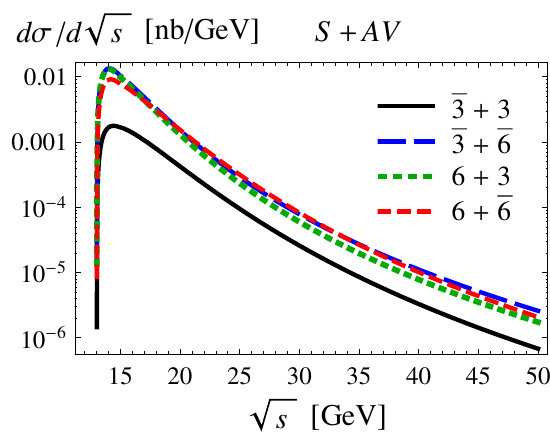}\includegraphics{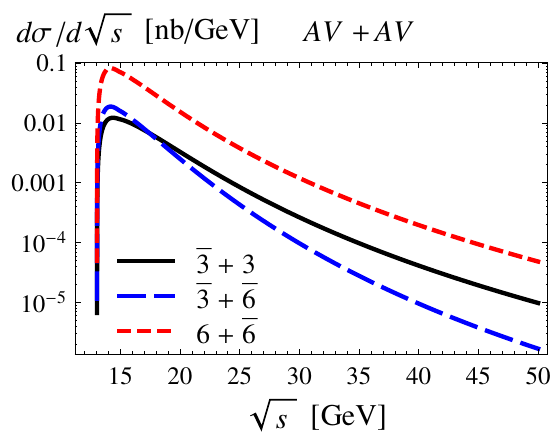}
\end{center}
\vspace{-.7cm}
\caption{The invariant mass $\sqrt s$ distributions for $S+S$ (left), $S+AV$ (center), and $AV+AV$ (right) pair double heavy diquark production at the energy $\sqrt S=7$~TeV.}
\label{fig:s-distr}
\end{figure}

Table~\ref{tbl1} contains numerical results for all 10 independent spin (scalar and axial-vector) and color (triplet and sextuplet) state combinations of the final diquarks contributing to $\sigma[pp\to D_{bc} + \bar D_{\bar b \bar c}+X]$.
The other 6 cross sections can be obtained by interchanging with simultaneous charge conjugation of the diquark pair, so that $\sigma[S{(bc)}_{6}+S{(\bar b\bar c)}_{3}]=\sigma[S{(bc)}_{\bar3}+S{(\bar b\bar c)}_{\bar 6}]$, $\sigma[AV{(bc)}_{6}+S{(\bar b\bar c)}_{3}]=\sigma[S{(bc)}_{\bar3}+AV{(\bar b\bar c)}_{\bar 6}]$ and so on. Formally, in the case of final diquark pair with unequal spin states, the diquark masses have to be interchanged too, but due to their very close proximity in the actual case of scalar and axial-vector diquarks ($M_{S(bc)}=6.517$~GeV and $M_{AV(bc)}=6.526$~GeV) the numerical effects on the cross section will be negligible. Table~\ref{tbl1} clearly shows that gluonic fusion sub-processes involving one or two final diquarks in a(an) (anti-)sextuplet color state provide considerable contributions, which basically prove to significantly exceed the corresponding triplet--antitriplet results. The largest contribution comes from the process of double axial-vector production, where the cross section of $6+\bar6$ pair is more than 5 times larger than its $\bar3+3$ counterpart, and the cross sections for two other possibilities $\bar3+\bar6$ and $6+3$ are both higher by almost 20\% in comparison with $\sigma[AV{(bc)}_{\bar3}+AV{(\bar b\bar c)}_{3}]$.
The sextuplet contributions are also important for $S+AV$ and $AV+S$ diquark spin pairs with all three additional sub-processes enhancing the purely triplet result by the factor of 4.5--5.5 each. Then, the simultaneous scalar and axial-vector diquark production acquires the same order of magnitude as the main $AV+AV$ channel, in spite of $\sigma[S{(bc)}_{\bar3}+AV{(\bar b\bar c)}_{3}]$ having the smallest value between all triplet--antitriplet states. The scalar--scalar diquark pair appears to be of the least significance, since its exclusively sextuplet case $6+\bar6$ has only 60\% of the double triplet $\bar3+3$ cross section, and both cases of mixed color combinations are even more suppressed by the factor of 9.
So, the $S+S$ diquark contributions to the pair double heavy baryon production are expected to be negligible.
The stated relations between the various sub-processes of $gg\to D_{bc} + \bar D_{\bar b \bar c}$ are independent of the choice of partonic functions $f_{g/p}(x)$ or collision energy $\sqrt S$, and they also remain valid with relativistic corrections taken into account.
Moreover, they demonstrate a remarkable stability through almost the whole range of $(bc)$ diquark rapidity $y$ for the corresponding differential cross section $d\sigma/dy$, as it showed in Fig.~\ref{fig:y-distr}. The invariant mass distributions presented in Fig.~\ref{fig:s-distr} have almost the same asymptotics at large values of $\sqrt s$ with the only exception of $\bar 3+\bar 6$ cross sections in $S+S$ and $AV+AV$ cases, which are suppressed by one additional power of $s^{-1/2}$ in comparison with the sextet--antisextet and antitriplet--triplet color pairs.

The cross section of pair diquark production in proton--proton interaction $\sigma[pp\to D_{bc} + \bar D_{\bar b \bar c}+X]$
gives an upper bound for the yield of the double heavy baryon pairs $(bc\,l_1)+(\bar b\bar c\,\bar l_2)$ in the same process summed over spin states of both baryons and also over all possible flavors of light quarks and antiquarks $l_{1,2}=u,~d,~s$. Adding up the contributions from Table~\ref{tbl1} for the collision energy $\sqrt S=7(14)$~TeV and CTEQ5L functions, we obtain the estimates 3.3(7.3)~nb and 0.9(2.1)~nb in nonrelativistic and relativistic cases, respectively, where the purely sextuplet channel is responsible for slightly more than a half of the result (1.7(3.8)~nb and 0.5(1.1)~nb), final states with mixed colors $\bar3+\bar6$ and $6+3$ contribute about 37\% of the cross section (1.2(2.7)~nb and 0.3(0.8)~nb), and the 12\% residue (0.4(0.8)~nb and 0.1(0.2)~nb) originates from the conventional triplet--antitriplet diquark pairs. So, under the assumption that both triplet and sextuplet states are described by the same value of $| R(0)|^2$, the processes involving one or two final sextuplets are the dominant ones for pair double heavy diquark and baryon production, and they will still remain to prevail over the triplet only cases even if the values of their nonperturbative parameters corresponding to $| R(0)|^2$ are found to be overestimated as mush as 3--4 times.

\begin{figure}[t!]
\vspace*{.7cm}
\begin{center}
\includegraphics{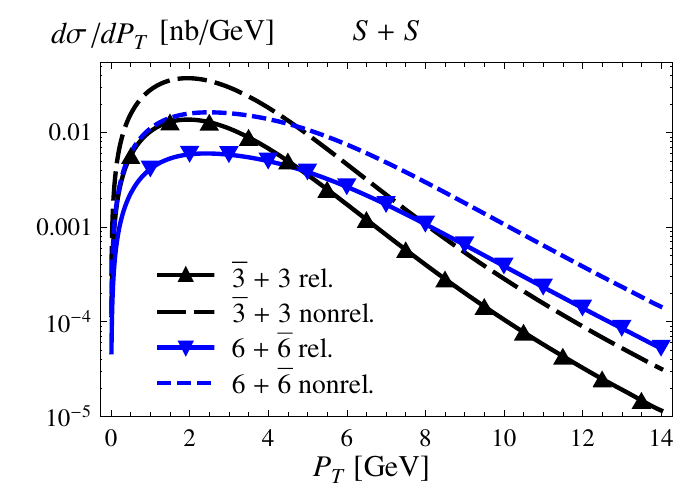}\quad\includegraphics{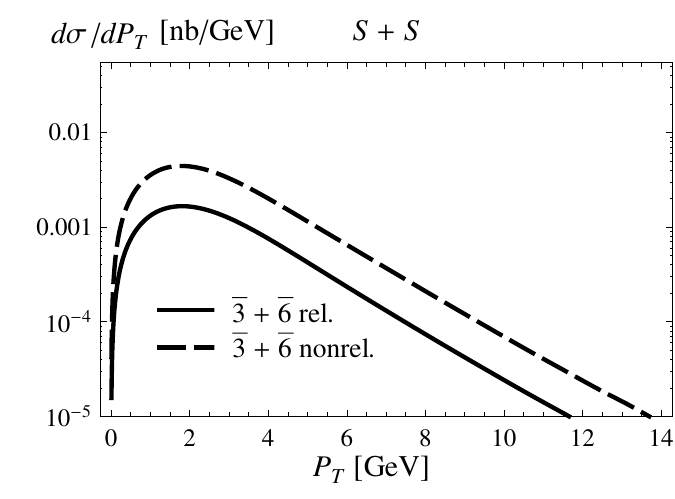}

\includegraphics{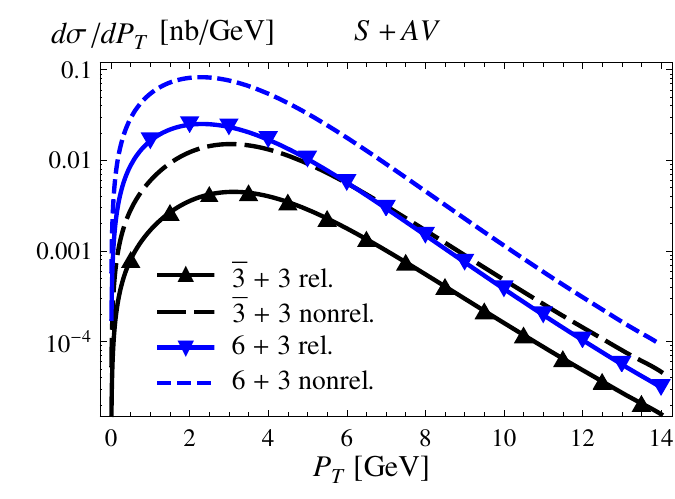}\quad\includegraphics{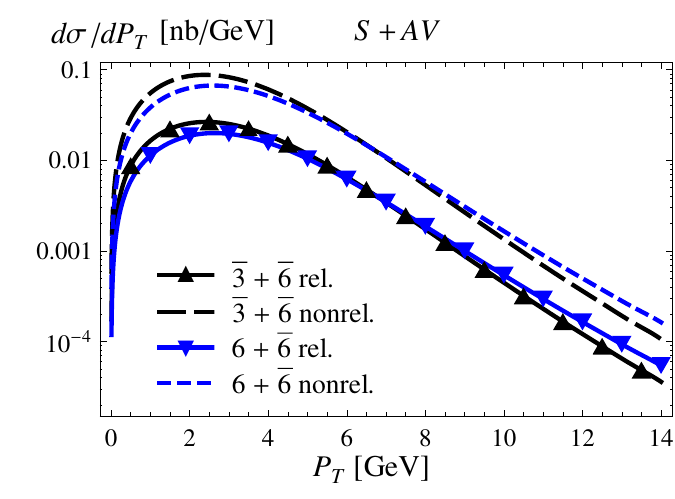}

\includegraphics{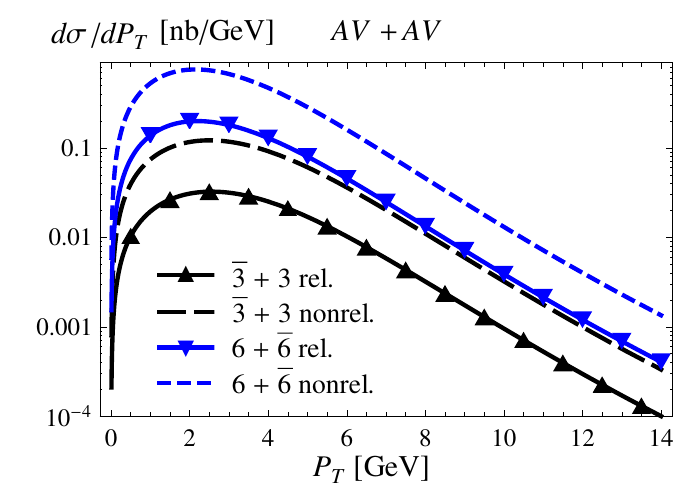}\quad\includegraphics{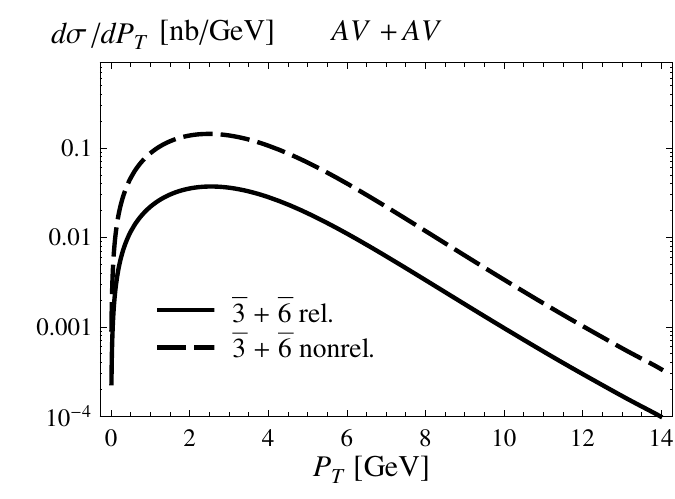}
\end{center}
\vspace{-.7cm}
\caption{The transverse momentum $P_T$ distributions for $S+S$ (upper panel), $S+AV$ (center), and $AV+AV$ (lower panel) pair double heavy diquark production at the energy $\sqrt S=14$~TeV. The results in nonrelativistic approximation (nonrel.) and with relativistic effects (rel.) are shown.}
\label{fig:pt-distr}
\end{figure}

The cross sections~\eqref{eq:cs} and~\eqref{eq:cs2} contain several types of relativistic corrections with the net effect of up to four times decreasing in comparison with the completely nonrelativistic analysis. Such tremendous fall is entirely caused by the Breit-like interaction between (anti)quarks in the bound state determining the masses and wave functions of scalar and axial-vector diquarks.
The transition from nonrelativistic parameter $|R^\text{NR}(0)|^4$ to its relativistic generalization $| R_{D_{bc}}(0)|^2| R_{\bar D_{\bar b \bar c}}(0)|^2$ suppresses the cross section by factor  2.3--2.9 depending on the final spin content. Note that direct corrections to the diquark wave functions induced by relativistic terms in quark--quark interaction potential are not so large. According to Table~\ref{tbl0}, the difference between $R^\text{NR}(0)$ and $ R(0)$ does not exceed 20\%, but its forth power actually entering expressions~\eqref{eq:cs} and~\eqref{eq:cs2} significantly amplifies the decreasing effect. The diquark masses enter cross sections~\eqref{eq:cs} and~\eqref{eq:cs2} as prefactors and also essentially influence the functions $F^{(i)}(s,t)$, so that the bound state corrections connected with the non-zero values of bound state energies $W_{S,AV(bc)} = M_{S,AV(bc)}- m_c -m_b\ne0$ can be identified. These corrections also turn out to be negative and decrease the cross sections by 20--30\% against the results calculated in nonrelativistic approximation $M_{S(bc)}=M_{AV(bc)}=m_b+m_c$.
Finally, our analysis includes perturbative corrections of the second order in (anti)quarks relative momenta $p$ and $q$ (or, alternatively, $\mathcal O(v^2)$ order in relative velocity $v$), which are directly determined by relativistic structure of the production amplitude~\eqref{eq:m-main}. As it was mentioned earlier, such type of corrections are represented by the terms $\omega_{nk}F^{(i)}(s,t)$ in Eqs.~\eqref{eq:cs} and~\eqref{eq:cs2}. They generally add 8--12\% to all of the considered production channels with exception of the following three color--spin combinations of the final diquark pair: $S_{\bar3}+S_{\bar 6}$ (25\%), $AV_{\bar3}+AV_{3}$ (5\%), and $AV_{6}+AV_{\bar6}$ (1.5\%). The above example of scalar--scalar pair shows that perturbative corrections can give considerable enhancements, which are, however, negligible in this particular case due to the lowest numerical importance of $gg\to S(bc)_{\bar3}+S(\bar b \bar c)_{6}$ sub-process. In average, $\mathcal O(v^2)$ corrections contribute slightly more than $6\%$ to the upper estimate of 0.9(2.1)~nb for the cross section of double heavy baryons at $\sqrt S=7(14)$~TeV.
The relativistic corrections do not change the visible shape of all examined cross section distributions, as it shown in Fig.~\ref{fig:pt-distr} on the example of transverse momentum distribution $d\sigma/d P_T$.

The total error of the numerical results for cross sections~\eqref{eq:cs} and~\eqref{eq:cs2} presented in Table~\ref{tbl1} is basically given by the main sources of relativistic corrections and can be estimated as 44\%. The 10\% uncertainty of wave functions calculated in the potential model is responsible for the main 40\% error coming from the forth power of $ R(0)$~\cite{mt2014}. The perturbative contributions from forth and higher orders of the amplitude expansion can be taken for 10\%, since the calculated second order corrections are already small. The another 15\% error source is introduced by the uncertainty of CTEQ5L and CTEQ6L1 partonic distribution functions~\cite{cteqs,mt2013}. In order to obtain the final value of 44\%, all mentioned uncertainties are summed in quadrature.

In this paper, the complete study of pair $(bc)$ diquarks production at the LHC center-of-mass energies $\sqrt S=7$ and 14~TeV have been performed in the leading order of gluonic fusion channel. Several types of relativistic improvements to the cross section have been implemented according to the framework of relativistic quark model. The corresponding nonperturbative parameters calculated in assumption of Breit-like interaction between (anti)quarks of the bound state are found to significantly decrease the cross section in comparison to the estimates based on the completely nonrelativistic Cornell model. The perturbative $\mathcal O(v^2)$ corrections originating from the production amplitude expansions can give moderate improvements, although they turn out to be irrelevant for the main sub-process $gg\to AV{(bc)}_{6}+AV{(\bar b\bar c)}_{\bar 6}$ determining almost the half of the final result.
The total error of the calculation is estimated on the level of $50\%$, which is mostly determined by the accuracy of the double heavy diquark wave functions obtained in the considered model.
It is shown that sextuplet color states of the final diquarks appear to dominate the total cross section. So, the pair double heavy baryon production can represent a good test for the color-sextuplet production mechanism.

\begin{acknowledgments}
The work was supported  by the ''Dynasty`` foundation  and by the Ministry of Education and Science of Russia
under SSAU Competitiveness Enhancement Program 2013--2020.
\end{acknowledgments}

\appendix
\section{The structure of supplementary material}
The electronic supplementary material to the paper consists of 84 textual files containing the explicit form of the functions $F^{(i)}(s,t)$ entering the cross sections~\eqref{eq:cs} and~\eqref{eq:cs2} for all 10 independent spin and color states combinations of the final diquark pair.
The functions are linearly expanded in diquark bound energies $W_{bc}=M_{bc}-m_c-m_b$ and  $W_{\bar b\bar c}=M_{\bar b\bar c}-m_c-m_b$:
\be
F^{(i)}(s,t)=F_0^{(i)}(s,t)+W_{bc}\,F_1^{(i)}(s,t)+W_{\bar b\bar c}\,F_2^{(i)}(s,t),
\ee
where the last term exists only for the $S+AV$ final states with two diquarks of unequal masses.
Then, for example, the file \verb"F[3S+6AV](1).0" contains the function $F_0^{(1)}(s,t)$ of $S{(bc)}_{\bar3}+AV{(\bar b\bar c)}_{\bar 6}$ pair.

\end{document}